\documentclass[9pt,sigconf,authorversion]{acmart-modified}
\PassOptionsToPackage{bookmarks=false}{hyperref}

\RequirePackage{epsfig} % good old way of including figures
\RequirePackage[small,compact]{titlesec} % squeezing whitespace between section heading

\newcommand{\eqnref}[1]{(\ref{eq:#1})}
\newcommand{\eqnlabel}[1]{\label{eq:#1}}
\newcommand{\figlabel}[1]{\label{fig:#1}}
\newcommand{\figref}[1]{Fig.~\ref{fig:#1}}

\newcommand{\tabref}[1]{Tab.~\ref{tab:#1}}
\newcommand{\tablabel}[1]{\label{tab:#1}}

\newcommand{\httpref}[1]{}

\newcommand{\aka}{{\it aka}}

\newcommand{\etc}{\textit{etc.\@}}                                                
\renewcommand{\jmath}{j}

\newcommand{\be}[1]{\begin{equation}\label{#1}}
\def\ee{\end{equation}}

\newcommand{\ignore}[1]{}

\newcommand{\COMMENT}[1]{}

\ignore{
    
}

	\titlespacing{\section}{0pt}{*0}{*0.1} % {0pt} is indentation; {*0} is 
	\titlespacing{\subsection}{0pt}{*0.1}{*0}
	\titlespacing{\subsubsection}{0pt}{*0.1}{*0}
\AtBeginDocument{%
  \providecommand\BibTeX{{%
    \normalfont B\kern-0.5em{\scshape i\kern-0.25em b}\kern-0.8em\TeX}}}

\copyrightyear{2019}
\acmYear{2019}
\setcopyright{acmcopyright}
\acmConference[DAC '19]{The 56th Annual Design Automation Conference 2019}{June 2--6, 2019}{Las Vegas, NV, USA}
\acmBooktitle{The 56th Annual Design Automation Conference 2019 (DAC '19), June 2--6, 2019, Las Vegas, NV, USA}
\acmPrice{15.00}
\acmDOI{10.1145/3316781.3322473}
\acmISBN{978-1-4503-6725-7/19/06}

\begin{document}

\author{Tianshi Wang}
\affiliation{%
  \institution{\vspace{-0.5em}University of California, Berkeley\vspace{-0.7em}}
  \city{Berkeley}
  \state{CA}
  \country{USA}}

\author{Leon Wu}
\affiliation{%
  \institution{\vspace{-0.5em}University of California, Berkeley\vspace{-0.7em}}
  \city{Berkeley}
  \state{CA}
  \country{USA}}

\author{Jaijeet Roychowdhury}
\affiliation{%
  \institution{\vspace{-0.5em}University of California, Berkeley\vspace{-0.7em}}
  \city{Berkeley}
  \state{CA}
  \country{USA}}

% \vspace{-2em}

\title{
Late Breaking Results: New Computational Results and Hardware Prototypes for Oscillator-based Ising Machines \vspace{-0.3em}}

\thispagestyle{empty}

\begin{abstract}
In this paper, we report new results on a novel Ising machine technology for solving combinatorial optimization problems using networks of coupled self-sustaining oscillators. Specifically, we present several working hardware prototypes using CMOS electronic oscillators, built on breadboards/perfboards and PCBs, implementing Ising machines consisting of up to 240 spins with programmable couplings. We also report that, just by simulating the differential equations of such Ising machines of larger sizes, good solutions can be achieved easily on benchmark optimization problems, demonstrating the effectiveness of oscillator-based Ising machines.
\end{abstract}

\renewcommand{\shortauthors}{Wang, et al.}
\settopmatter{printacmref=false}

\maketitle

\section{Introduction}
The Ising model is a mathematical model originally used to study
ferromagnetism.
It describes spins coupled in a graph that try to minimize a collective
energy, \aka, the Ising Hamiltonian.
\vspace{-0.5em}
\begin{equation}\eqnlabel{IsingH}
%\begin{align}
    \min ~~ H \triangleq - \sum_{1 \leq i<j \leq n} J_{ij} s_i s_j - \sum_{i=1}^n h_i s_i, 
    \quad \text{s.t.} ~ s_i \in \{-1,~+1\},
%\end{align}
\vspace{-0.4em}
\end{equation}
where $n$ is the number of spins; $\{J_{ij}\}$ and $\{h_i\}$ are real coefficients.

Finding optimal spin configurations that minimize the Ising Hamiltonian, \aka,
the Ising problem, is in general difficult
\cite{barahona1982computational}, even with purposely built digital
accelerators \cite{yamaoka2016IsingCMOS}.
A physical implementation of coupled spins that directly perform the
minimization in an analog way, namely an Ising machine, therefore becomes very
attractive for potential speed and power advantages.
As many difficult real-world optimization problems are equivalent to the
Ising problem \cite{lucas2013ising}, Ising machines have been attracting
considerable research attention in recent years, with incarnations mostly based
on novel devices, such as optical cavities \cite{inagaki2016ScienceIsing2000},
nanomagnets \cite{camsari2017pbits}, and quantum circuits \cite{bian2014Ising}.
A recent work --- oscillator-based Ising machine (OIM)
\cite{WaRoOscIsing2017,WaRo2019OIMarXiv} --- shows that almost all types of
nonlinear self-sustaining oscillators are suitable to represent Ising spins
physically.
As many tried-and-tested types of such oscillators already exist, this scheme
offers the advantages of scalability to large numbers of spins, high-speed and
low-power operation, and straightforward design and fabrication using standard
circuit technologies.

In the recent months, we have achieved several results from testing the feasibility
of the OIM idea:
\newcounter{listcounter}
\begin{list}{$\arabic{listcounter}$. }
{
	\usecounter{listcounter}
	\setlength{\topsep}{1pt}
	\setlength{\leftmargin}{8pt}
	\setlength{\labelsep}{1pt}
	\setlength{\rightmargin}{0pt}
	\setlength{\labelwidth}{10pt}
    \setlength{\itemsep}{0pt}
}
\item \textbf{We have built several hardware prototypes}, starting from coupling 8 CMOS
LC oscillators on a breadboard, moving on to soldering 32 of them on perfboards, then to
PCB designs of size 64 and 240 with programmable couplings. We have tested all
prototypes on many instances of Ising problems of their corresponding sizes;
every prototype can achieve global optima for these problems.
We plan to make them \textbf{open hardware projects}, with schematics, design
files, tutorials and code released to the public, so that researchers (and
hobbyists) can reproduce the results and improve upon the design.

\item We have simulated larger-sized OIMs, trying them on \textbf{all 54
problems in the G-set} \cite{helmberg2000spectral} (available at \cite{GsetWebsite})
--- a widely used benchmark set for MAX-CUT problems (which have a
direct mapping to Ising problems).
Much to our surprise, without changing any parameters (coupling strength, noise
level, \etc) across different problems, the results match 21/54 and improve upon 17/54
previously published optimal solutions.
% Our plan is to open-source release the code (a simple C++ script for
% integrating stochastic differential equations), together with the random seeds
% used for generating these results.
\end{list}
The remainder of this paper contains a brief overview of OIM's mechanism and
more details on the new results listed above.

\section{Oscillator-based Ising Machines}

OIM's operation relies on a special type of injection locking, known as 
Subharmonic Injection Locking (SHIL).
Under SHIL, when an oscillator is perturbed by an input at twice its natural
frequency (often called a SYNC signal), it can develop bistable phase-locked
states, separated by a $180^\circ$ phase difference.

\begin{figure}[htbp]
	\vspace{-0.3em}
	\centering
	{
		\epsfig{file=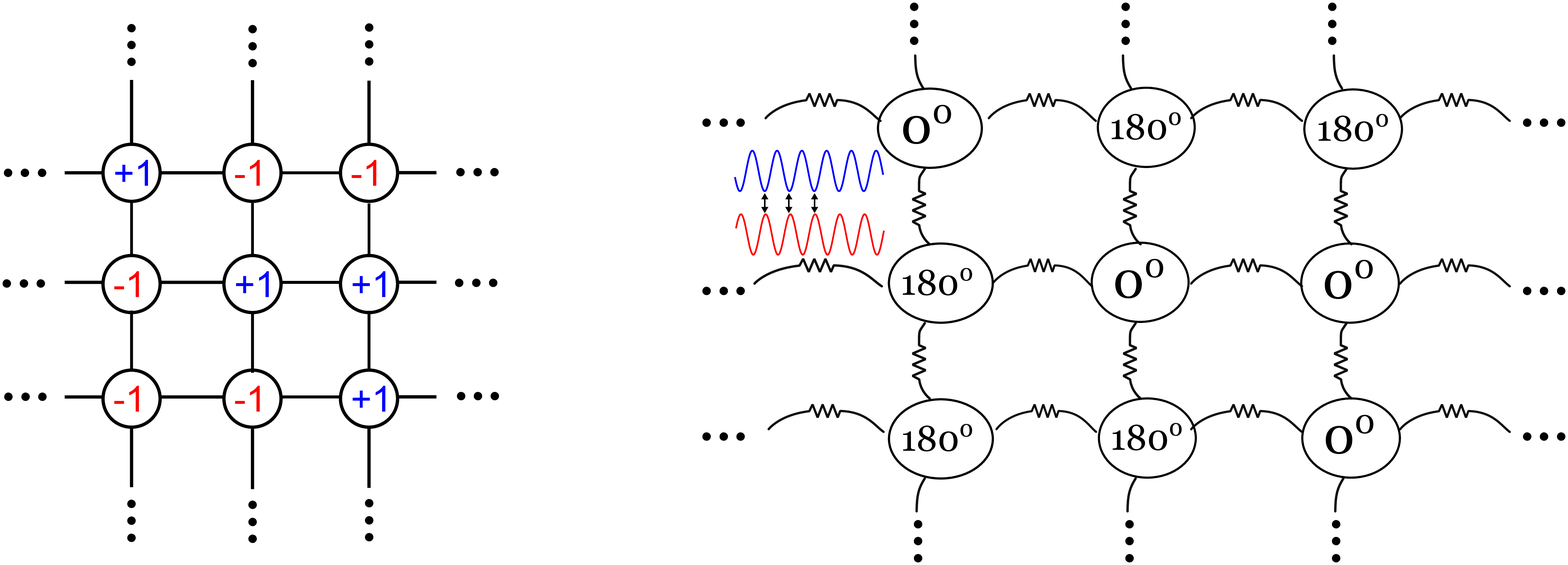, width=0.9\linewidth}
	}
    \caption{An Ising model and its OIM schematic.\figlabel{fig1}}
\end{figure}

Multiple such oscillators, whose phase is ``binarized'' by a common SYNC
through SHIL, can be networked such that they synchronize to binary phase
configurations that minimize Ising Hamiltonians, as illustrated in
\figref{fig1}.
It can be shown \cite{WaRoOscIsing2017} that when the coupling coefficients (often
represented by conductances connecting electronic oscillators\footnote{\scriptsize A negative coefficient
can be implemented conveniently using positive conductances to cross-couple two differential oscillators.}) are proportional to
$J_{ij}$s and $h_{i}$s in \eqnref{IsingH}, the oscillator network naturally
minimizes a global Lyapunov function that can be made equivalent to the Ising
Hamiltonian, thus physically implementing an Ising machine.

\section{Hardware Prototypes}

\begin{figure*}[htbp]
	\centering
	{
		\hspace{0em}\epsfig{file=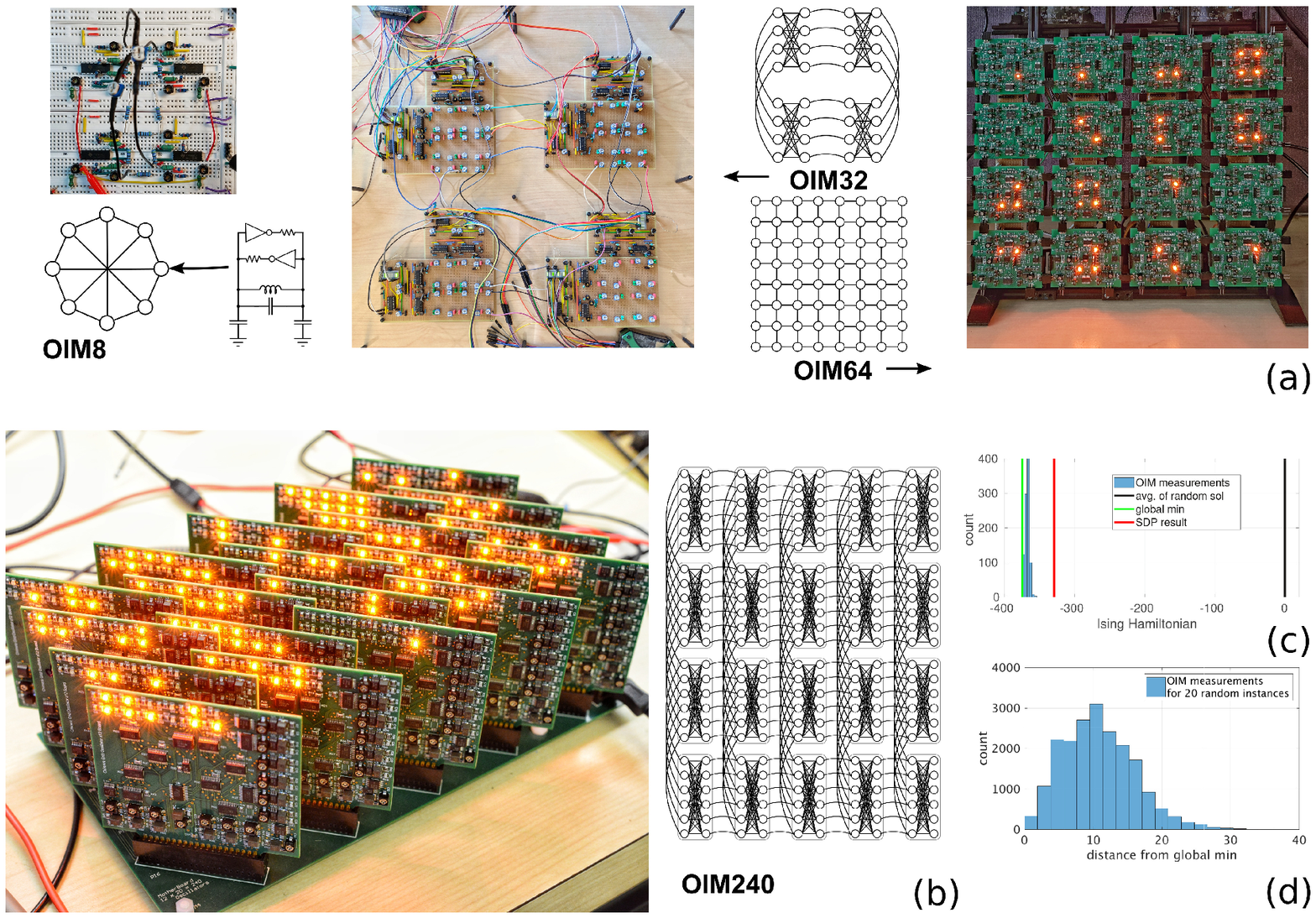, width=0.98\linewidth}
	}
	% \vspace{-0.5em}
    \caption{OIM prototypes: (a) photos and schematics of OIM8, OIM32, OIM64; (b) OIM240;
(c) energy levels of 1000 measured solutions from OIM240, on a random instance of size-240 Ising problem;
(d) energy levels for 20 such random instances (showing the distances to their respective global minimum).\figlabel{Ising_hardware_history} \vspace{-2.0em}}
	
\end{figure*}

The schematics of several OIM hardware prototypes are summarized in
\figref{Ising_hardware_history}.
They all use CMOS LC oscillators made with cross-coupled inverters (from TI
SN74HC04N ICs), fixed inductors, trimmer capacitors and a 5V single supply.
OIM8 and OIM32 use 33$\mu$H inductors with capacitors tuned to around 30pF, for
a natural frequency of 5MHz.
We could manually plug in resistors and potentiometers on the breadboard to try
different problems on OIM8, then read back results using oscilloscopes.
For OIM32, rotary potentiometers were soldered on perfboards as couplings.
Next to each potentiometer, we designed male pin connectors such that the
polarity of each connection can be controlled by shorting different pins using
female jumper caps (color coded green and pink for positive and negative
couplings).
Furthermore, we soldered TI SN74HC86N Exclusive-OR (XOR) gate ICs to convert the
oscillator phases to voltage levels, which then power on-board LEDs for
visualization and readout (by two 16-channel logic analyzers).
We observed that for small-sized (8 and 32) Ising problems, global optima can
be achieved easily using these prototypes.

OIM64 and OIM240 use digital potentiometers from AD5206 ICs (6-channel
potentiometers with 8-bit accuracy).
Because these ICs are designed primarily for audio processing and do not have
multi-MHz bandwidth, we reduced oscillator frequency to 1MHz.
Both prototypes consist of multiple PCBs.
OIM64 connects 64 oscillators in a 8x8 2D toroidal grid, with 192 couplings,
each made of one channel of AD5206 and a SPDT switch for setting its polarity.
Even though it was not easy to ``program'' OIM64 due to the use of physical
switches, we tried it on 10 randomly generated toroidal Ising grid instances,
achieving the global optimum for each one.

In OIM240, we improved the design to use the position of the potentiometer
wiper to switch polarity, thus eliminating the use of switches and making the
coupling truly programmable.
On each PCB, we implemented 12 oscillators with a denser connectivity; 20 such
PCBs were plugged into a motherboard through edge connectors, and
interconnected in a 4x5 toroidal grid, implementing a total of 240 oscillators
with 1200 couplings.
The motherboard also distributes CLK, data lines, address lines for programming
the 200 AD5206 ICs and for reading oscillator states, all controlled by an
Arduino module on the motherboard that communicates with a PC through
USB.
When operating OIM240, we flip on the supply digitally, wait 1ms for
oscillators to synchronize, then read back the solution.
Even with all the overhead from serial reading, solutions can be read back every
3.5ms.
OIM240's operation consumes $\sim$5W of power for all the oscillators and
peripheral circuitry, excluding only the LEDs.

We tested OIM240 with many randomly generated Ising problems (with each of the 1200
couplings randomly chosen from 0, $-1$, $+1$).
A typical histogram for the energy levels of the measured solutions is shown in
\figref{Ising_hardware_history} (c).
Note that a random (trivial) solution has an energy around 0, whereas the best
polynomial-time algorithm (based on SDP) guarantees to achieve 87.8\% of the
global optimum.
In comparison, results from OIM240 center around a very low energy, and achieve
the global optimum multiple times.
We performed the same measurements for 20 different random Ising problems, with
the distances of solutions from their respective global
optima\footnote{\scriptsize We ran simulated annealing for a long time (1min)
and for multiple times, then treated the best results as global optima.} shown
in \figref{Ising_hardware_history} (d).
The fact that OIM240 is finding highly non-trivial solutions indicates that it
indeed physically implements a working Ising machine.

\section{Results on MAX-CUT Benchmark Problems}
We simulated OIMs for solving the Ising version of the MAX-CUT problems from
the G-set, with sizes ranging from 800 to 3000, with partial results shown in
\tabref{results}. Full results for all 54 problems will be open-source released
with the code (a simple C++ script for integrating stochastic differential
equations).
Each problem was simulated with 100 random instances.
In \tabref{results}, we compare the results with the best-known results listed
in \cite{GsetWebsite} (mostly from the heuristic algorithm Scatter Search (SS)
\cite{marti2009SS}). We also list results from a recent study applying
simulated annealing to MAX-CUT \cite{myklebust2015SA}, the only one we could
find that contains results for all the G-set problems.
Although we could not rerun the best-known results reported in
\cite{GsetWebsite}, we made the comparison more fair by running our simulation
with an older desktop that are comparable with the environment used by others.

\begin{table}[ht!]
    \begin{center}
% {\footnotesize
{\scriptsize
        \begin{tabular}{|c|cc|cc|cc|cc|ccccc|}
        \hline
{\bf Benchmark} & {\bf SS} & {\bf Time (s)} & {\bf SA} & {\bf Time (s)} & {\bf OIM} & {\bf Time (s)} \\
        \hline
G1 & {\bf 11624} & 139 & 11621 & 295 & {\bf 11624} & 52.6  \\ \hline
G11 & 562 & 172 & {\bf 564} & 189 & {\bf 564} & 6.7 \\ \hline
G21 & 930 & 233 & 927 & 195 & {\bf 931} & 14.6 \\ \hline
G31 & 3288 & 1336 & {\bf 3309} & 214 & 3301 & 59.1 \\ \hline
G41 & 2386 & 1017 & {\bf 2405} & 208 & 2401 & 37.8 \\ \hline
G51 & {\bf 3846} & 513 & 3841 & 234 & {\bf 3846} & 18.4 \\ \hline
        \end{tabular}
}
    \end{center}
  \caption{\scriptsize Results (cut sizes) from OIM and several heuristics run on MAX-CUT benchmarks in the G-set. Time reported in this table is for a single run.\tablabel{results}}
\end{table}

Note that we were simulating a fixed duration equivalent to 1000 cycles of
oscillation, which will be much faster on a physical hardware than the
simulation time we show here.
But even simulating OIM's differential equations yields a good solution
quality.
Unlike other algorithms, OIM simulation does not know about the energy function
or relative energy changes, which are implicit in the dynamics of differential
equations, yet it proves effective and fast.

\begin{figure}[htbp]
	\centering
	{
		\epsfig{file=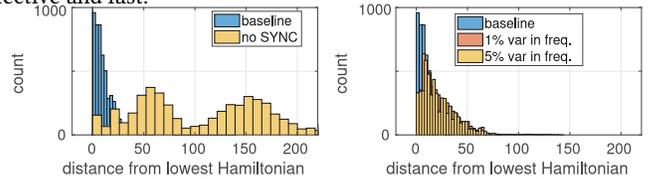, width=0.89\linewidth}
	}
	\vspace{0.2em}
    \caption{Hamiltonian values achieved by several variants of OIM.\figlabel{hist2}}
\end{figure}

Furthermore, we removed SYNC from OIM, reran the 54 benchmarks and compared the results in \figref{hist2}.
Without SYNC, the system becomes a simple coupled oscillator system with phases
taking a continuum of values (as opposed to binary values), which we then
threshold as Ising solutions.
From \figref{hist2}, the solutions become much worse, indicating SYNC and SHIL
are essential to OIM's operation.
We have also studied OIM's performance with variability in the natural
frequency of oscillators (a spread following a Gaussian distribution of 1\% or
5\% standard derivation).
From \figref{hist2}, we observe that even with non-trivial variability, the
solution quality is not affected by much.

\section{Conclusion and Outlook}
While much future work is still needed to establish OIM's advantages or
competence against other Ising machine proposals, the new preliminary results
we show in this paper demonstrate that OIM is indeed a feasible and attractive
Ising machine technology.
Even with PCB implementations using decade-old 7404 inverters, 1MHz oscillators
and 5V supply, OIM can return solutions in milliseconds with only 5W of power
consumption.
Future CMOS IC implementations at a larger scale with GHz nano-oscillators
under lower supply voltages hold exciting promises in outperforming
conventional hardware for Ising problems; our results shown here will server as
a solid foundation for these future developments.

\thispagestyle{empty}

\let\em=\it
\bibliographystyle{unsrt}
{\tiny\bibliography{von-Neumann-jr,PHLOGON-jr,jr,tianshi}}

\thispagestyle{empty}

\end{document}